\documentclass[aip,reprint]{revtex4-1}

\usepackage{color}
\usepackage{graphicx}
\usepackage{amsmath}
\usepackage{subcaption}

\usepackage{natbib}

\begin{document}
\title{Updated Magnetized Transport Coefficients: Impact on Laser-Plasmas with Self-Generated or Applied Magnetic Fields}

\author{C. A. Walsh}
\affiliation{Lawrence Livermore National Laboratory}
\author{J. D. Sadler}
\affiliation{Los Alamos National Laboratory}
\author{J. R. Davies}
\affiliation{Laboratory for Laser Energetics, University of Rochester}
\email{walsh34@llnl.gov}

\date{\today}

	\begin{abstract}
		Errors in the Epperlein \& Haines [PoF (1986)] transport coefficients were recently found at low electron magnetizations, with new magnetic transport coefficients proposed simultaneously by two teams [Sadler, Walsh \& Li, PRL (2021) and Davies, Wen, Ji \& Held, PoP (2021)]; these two separate sets of updated coefficients are shown in this paper to be in agreement. The importance of these new coefficients in laser-plasmas with either self-generated or applied magnetic fields is demonstrated. When an external magnetic field is applied, the cross-gradient-Nernst term twists the field structure; this twisting is reduced by the new coefficients in the low magnetization regime. For plasmas where only self-generated magnetic fields are present, the new coefficients are found to result in the magnetic field moving with the Righi-Leduc heat-flow, enhancing the impact of MHD. Simulations of Biermann Battery magnetic fields around ICF hot-spot perturbations are presented, with cross-gradient-Nernst transport increasing spike penetration.
	\end{abstract}
	\maketitle
	
	\section{Introduction}

	Magnetic fields exist in all plasmas: in laser-driven ablation fronts \cite{gao2015,campbell2020,manuel2012}, ICF capsules \cite{igumenshchev2014,hill2017,walsh2017}, shock fronts \cite{PhysRevLett.123.055002}, hohlraums \cite{farmer2017,sherlock2020} and Z-pinches \cite{gomez2020}. Magnetic fields can also be purposefully applied to ICF implosions to improve fusion performance \cite{chang2011,perkins2017,walsh2019,slutz2010}. The transport of these magnetic fields within the plasma is critically important to the design and interpretation of experiments. Non-local VFP simulations capture the transport processes innately by resolving the distribution of electron energies \cite{sherlock2020,hill2017,joglekar2014}. However, these calculations are prohibitively expensive and unstable, so simplified models with prescribed transport coefficients are used for spatial and temporal scales of interest. These transport coefficients are calculated by fitting to VFP data. The seminal work in this area is that of Braginskii, who laid the groundwork for the anisotropic transport of thermal energy and magnetic fields within the magneto-hydrodynamic (MHD) framework \cite{braginskii1965}. Following on from Braginskii, Epperlein \& Haines improved the transport coefficients, allowing for closer agreement to VFP calculations \cite{epperlein1986}. For 35 years the transport coefficients of Epperlein \& Haines were the agreed-upon standard to be implemented into extended-MHD codes \cite{walsh2017,farmer2017,perkins2017,watkins2018,igumenshchev2014}.
	
	\begin{figure}
		\centering
		\includegraphics[width=0.5\textwidth]{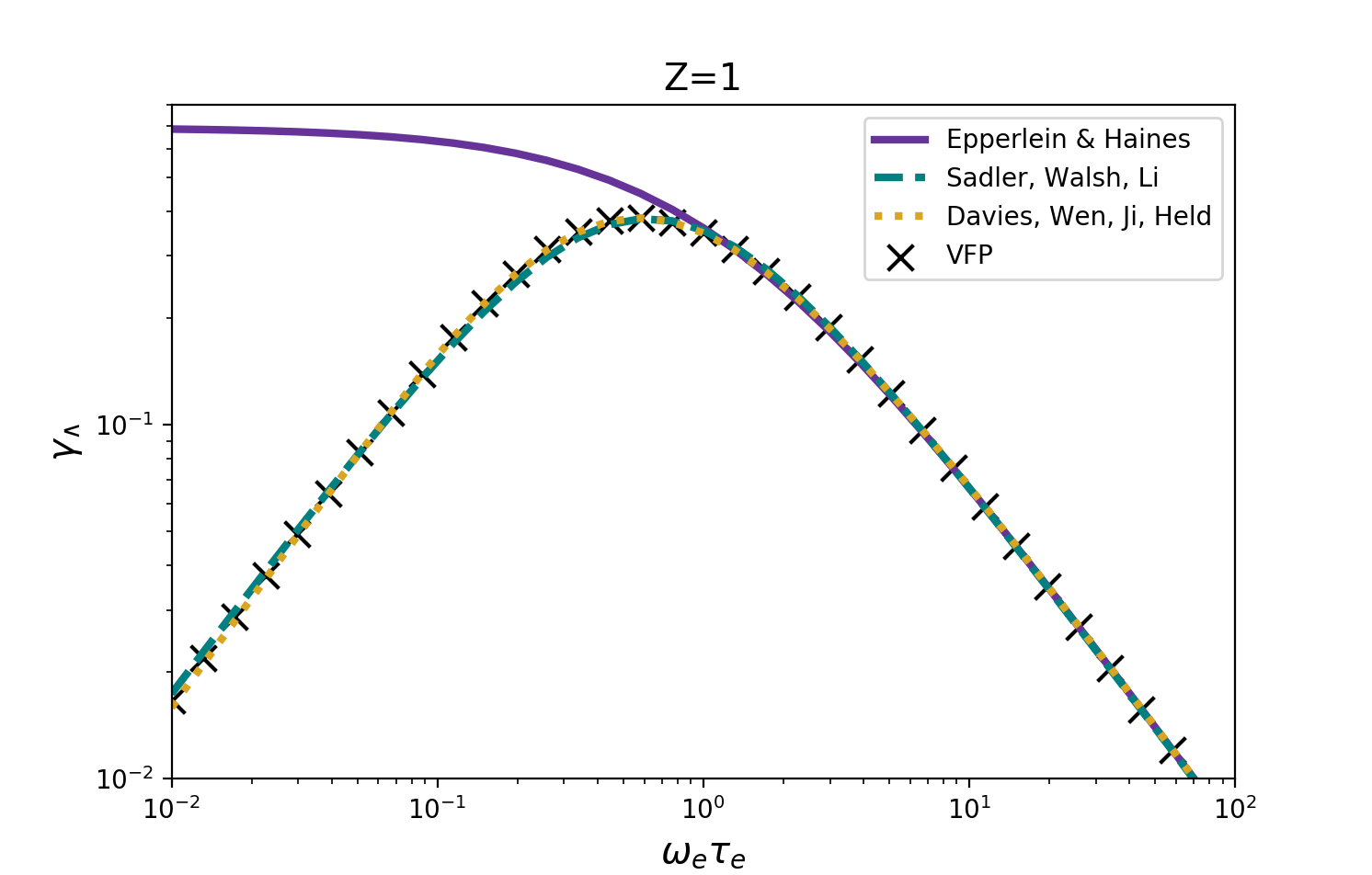}
		\caption{The cross-gradient-Nernst coefficient as a function of Hall Parameter calculated using different sources. All coefficients are attempting to replicate those given by VFP calculations. \label{fig:gamma}}
	\end{figure}
	
	Recently, simultaneous work by two research teams demonstrated errors in two of the transport coefficients at low electron magnetization \cite{sadler2021,davies2021}. While Epperlein \& Haines stated low errors in fitting the transport coefficients to VFP data, they did not consider that their transport coefficients were ill-formed. From their perspective the magnetic field evolution was captured by:
	
	\begin{align}
	\frac{\partial\underline{B}}{\partial{}t}=\nabla\times(\underline{v}\times\underline{B})-\nabla\times\frac{\underline{j}\times\underline{B}}{n_{e}e}-\nabla\times\frac{\underline{\underline{\alpha}}\cdot\underline{j}}{n_{e}^{2}e^{2}}\nonumber \\
	+\nabla\times\frac{\underline{\underline{\beta}}\cdot\nabla{}T_{e}}{e}+ \nabla \times \frac{\nabla P_{e}}{n_{e} e}\label{eq:mag_transport_old}
	\end{align}
	
	where $\underline{\underline{\alpha}}$ and $\underline{\underline{\beta}}$ are tensor transport coefficients calculated by fitting to 0D VFP simulations. While accurate, equation \ref{eq:mag_transport_old} does not make clear the impact of each term on the movement or generation of magnetic fields. Recent work re-arranged this equation to make the different physical processes more obvious\cite{walsh2020,davies2015,sadler2020}: 
	              
	\begin{align}
	\begin{split}
	\frac{\partial \underline{B}}{\partial t} = & - \nabla \times \frac{\alpha_{\parallel}}{\mu_0 e^2 n_e ^2} \nabla \times \underline{B} + \nabla \times (\underline{v}_B \times \underline{B} ) \\
	&+ \nabla \times \Bigg( \frac{\nabla P_e}{e n_e} - \frac{\beta_{\parallel} \nabla T_e}{e}\Bigg) \label{eq:mag_trans_new}
	\end{split}
	\end{align}
	where the first term is diffusive, the second term is advection of magnetic field at a velocity $\underline{v}_B$ and the final term is a source of magnetic flux. The advection velocity is then a combination of the bulk plasma motion, thermally-driven terms and current-driven terms:
	
	\begin{equation}
	\underline{v}_B = \underline{v} - \gamma_{\bot} \nabla T_e - \gamma_{\wedge}(\underline{\hat{b}} \times \nabla T_e) - \frac{\underline{j}}{e n_e}(1 + \delta_{\bot}^c) + \frac{\delta_{\wedge}^c}{e n_e} (\underline{j} \times \underline{\hat{b}})   \label{eq:mag_trans_new_velocity}
	\end{equation}
	
	where the transport coefficients have been re-written from the form given by Braginskii. The dimensionless forms (denoted by a super-script $^c$) depend only on Hall Parameter and effective Z~\cite{walsh2020}:
	
	\begin{equation}
		\gamma_{\bot}^c = \frac{\beta_{\wedge}^c}{\omega_e \tau_e}
	\end{equation}
	
	\begin{equation}
		\gamma_{\wedge}^c = \frac{\beta_{\parallel}^c - \beta_{\bot}^c}{\omega_e \tau_e}
	\end{equation}
	
	\begin{equation}
		\delta_{\bot}^c = \frac{\alpha_{\wedge}^c}{\omega_e \tau_e}
	\end{equation}
	
	\begin{equation}
		\delta_{\wedge}^c = \frac{\alpha_{\bot}^c - \alpha_{\parallel}^c}{\omega_e \tau_e}
	\end{equation}
	
	The dimensional forms of the thermally-driven coefficients are given by:
	\begin{equation}
	\gamma = \gamma^c \frac{\tau_e}{m_e}
	\end{equation}
	
	Once the transport coefficients have been re-written into this physically-motivated form, the fits of Epperlein \& Haines to the VFP simulations have large errors in $\gamma_{\wedge}^c$ and $\delta_{\wedge}^c$ at small $\omega_e \tau_e$. While the errors in fitting to $\beta_{\bot}^c$ are small, the $\gamma_{\wedge}$ numerator ($\beta_{\parallel}^c - \beta_{\bot}^c$) and denominator ($\omega_e \tau_e$) both tend to zero for small $\omega_e \tau_e$. This resulted in both $\gamma_{\wedge}^c$ and $\delta_{\wedge}^c$ being maximum for small $\omega_e \tau_e$, while the VFP simulations calculated them as exactly zero. Figure \ref{fig:gamma} shows how $\gamma_{\wedge}^c$ calculated using the values from Epperlein \& Haines deviate from the desired VFP data.
	
	New fits to the VFP data incorporate the knowledge of how the transport coefficients should be formulated to capture their limiting behavior at low Hall Parameter \cite{sadler2021,davies2021}. It should be reiterated that this newer work did not modify the VFP calculations, only changing how the transport coefficients were fit to the VFP data. Figure \ref{fig:gamma} includes $\gamma_{\wedge}^c$ calculated by Sadler, Walsh \& Li as well as by Davies, Wen, Ji \& Held; note that these two references give near-identical results. 
	
	\begin{figure}
		\centering
		\includegraphics[width=0.5\textwidth]{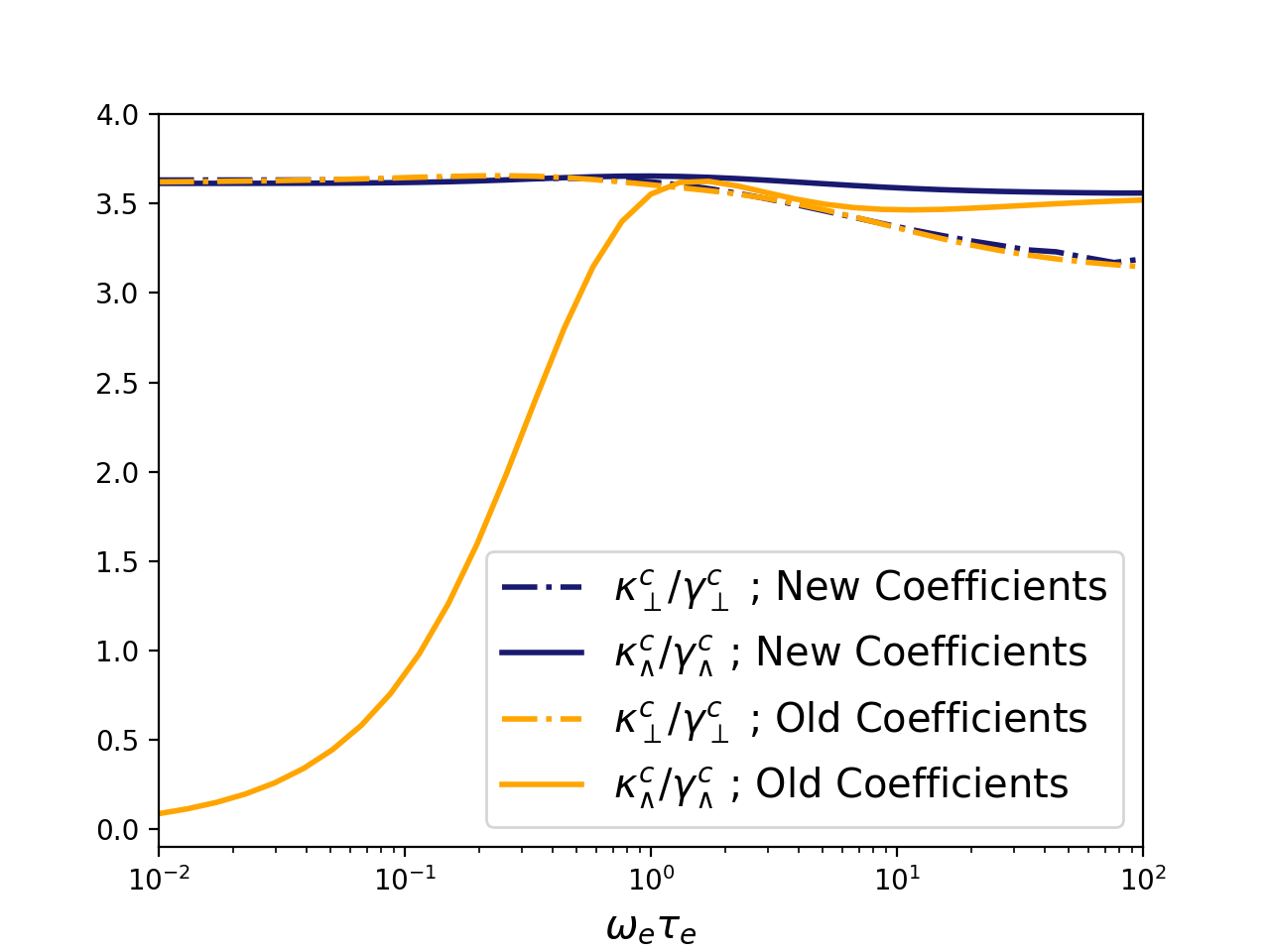}
		\caption{Ratio of thermal to magnetic transport coefficients as a function of electron Hall Parameter for the old and new coefficients. These are plotted for $Z=1$.\label{fig:coeff}}
	\end{figure}
	
	The updated transport coefficients make clear the similarities between thermal and magnetic transport. Looking specifically at the transport due to temperature gradients:
	
	\begin{align}
	&\underline{v}_{N} &= & &  -\gamma_{\bot}\nabla_{\bot} T_e  &- \gamma_{\wedge} \underline{\hat{b}} \times \nabla_{} T_e \label{eq:thermally_driven_transport1}\\
	&\underline{q}_{\kappa} &= &-\kappa_{\parallel} \nabla_{\parallel} T_e  &-\kappa_{\bot}\nabla_{\bot} T_e  &- \kappa_{\wedge} \underline{\hat{b}} \times \nabla_{} T_e \label{eq:thermally_driven_transport2}
	\end{align}
	
	where equation \ref{eq:thermally_driven_transport1} is magnetic transport and equation \ref{eq:thermally_driven_transport2} is thermal conduction. These equations are split into components parallel ($\parallel$) to the magnetic field, perpendicular to the field ($\bot$) and perpendicular to both the magnetic field and temperature gradient ($\wedge$). In equation \ref{eq:thermally_driven_transport1} the $\bot$ term is Nernst advection and the $\wedge$ term is cross-gradient-Nernst. For thermal conduction the terms are the unrestricted thermal conduction along field lines, suppressed thermal conduction perpendicular to the field and the Righi-Leduc heat-flow. 
	
	The analogy between Nernst ($\gamma_{\bot}^c$) and perpendicular heat-flow ($\kappa_{\bot}^c$) was originally noted by Haines \cite{haines1986}. Haines found that if the kinetic collision operator uses an artificial $v^{-2}$ dependence, where $v$ is the electron speed, there is an exact equivalence between heat flow and magnetic field motion, such that $\kappa_\perp^c/\gamma_\perp^c = \kappa_\wedge^c/\gamma_\wedge^c = 5/2$. The physical interpretation is that magnetic field is only frozen into the faster electrons in the distribution, since it can easily diffuse through the slower, more collisional electrons. The field therefore follows the faster electrons, which tend to move down temperature gradients with the heat flow. For the artificial collision operator, this intuition holds exactly for all $Z$ and $\omega_e\tau_e$. 
	
	However, this physical picture is unaffected by changing the $v$ dependence of the collision operator, so the character of magnetic field movement with the electron heat flow should still approximately hold. For example, if instead the collisions use the realistic Fokker-Planck $v^{-3}$ dependence, magnetic field should still move with the heat flow, although they are no longer directly proportional. With the Fokker-Planck collision operator, the ratios are no longer fixed at $5/2$, and they have dependence on $Z$ and $\omega_e\tau_e$. However, for fixed $Z$, the ratios remain approximately constant. Figure \ref{fig:coeff} shows the ratio $\kappa_{\bot}^c/\gamma_{\bot}^c$ against Hall Parameter for $Z=1$; this result is similar using the old or new coefficients, giving little variation.
	
	Figure \ref{fig:coeff} also plots the ratio $\kappa_{\wedge}^c/\gamma_{\wedge}^c$ against Hall Parameter for $Z=1$. Using the old Epperlein \& Haines coefficients the Righi-Leduc ($\kappa_{\wedge}^c$) coefficient is much smaller than cross-gradient-Nernst ($\gamma_{\wedge}^c$) at low magnetization. With the new transport coefficients, however, it can be seen that the cross-gradient-Nernst term is analogous to advection of magnetic field with the Righi-Leduc heat-flow; for $Z=1$, the ratio $\kappa_{\wedge}^c/\gamma_{\wedge}^c$ varies by less than 3\% across magnetization. 
	
	\begin{figure*}
		\centering
		\includegraphics[scale=0.8]{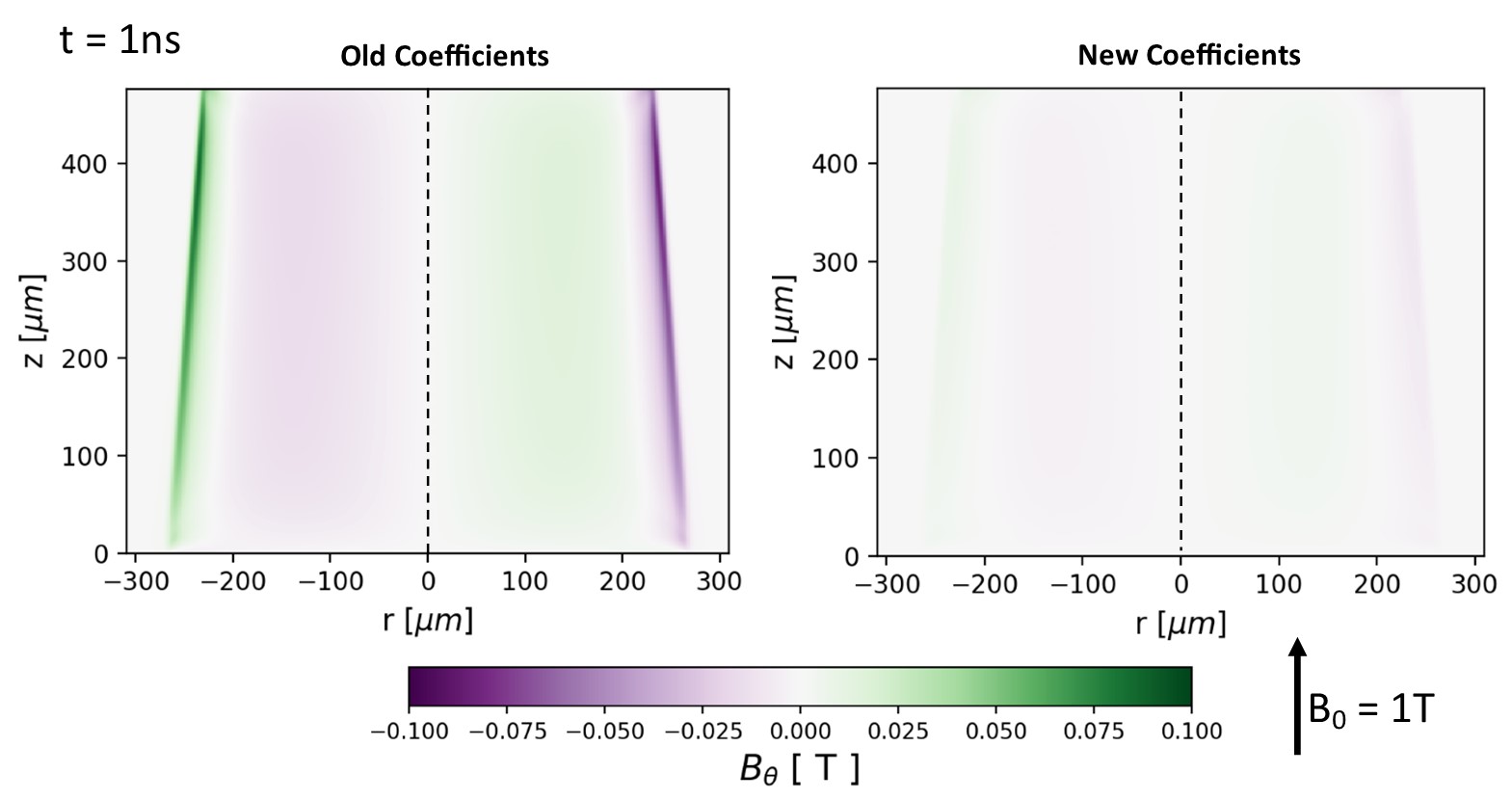}
		\caption{ \label{fig:underdense} Azimuthal magnetic field component generated by cross-gradient-Nernst advection for an under-dense magnetized plasma \cite{walsh2020} using the Epperlein \& Haines transport coefficients \cite{epperlein1986} (left) and the updated coefficients (right) \cite{sadler2021,davies2021}. These profiles are 1.0ns after the laser turns on. }
	\end{figure*}
	
	The ratio of thermal to magnetic transport coefficients does vary with plasma ionization. For low Hall Parameter, $\kappa_{\bot}^c/\gamma_{\bot}^c$ ranges between 3.63 for $Z=1$ to 1.39 for $Z=100$. This ratio is fundamental for systems where thermal conduction drives plasma ablation, where equation \ref{eq:thermally_driven_transport1} will move magnetic fields into the colder regions, while ablation driven by equation \ref{eq:thermally_driven_transport2} will counter-act that transport \cite{betti2001}. The balance of Nernst demagnetization to plasma ablation has been found to be critically important in magnetized plasma conditions \cite{manuel2015,hill2017,walsh2017,farmer2017,campbell2020,sherlock2020,walsh2020a,hill2021}. The decrease of $\kappa_{\bot}^c/\gamma_{\bot}^c$ with Z suggests that low-Z plasmas will be less impacted by Nernst de-magnetization than high-Z plasmas.
	
	The new fits make intuitive sense. The $\wedge$ terms come from electrons having curved orbits; if the magnetization is low, the electron trajectories should be straight on average. The Epperlein \& Haines fits captured this behavior for the heat-flow ($\kappa_{\wedge}$) but not for the magnetic transport ($\gamma_{\wedge}$). At large magnetization the electrons go through many orbits before colliding, meaning that there is no preferential direction for transport; this is captured in both the old and new coefficients, with $\kappa_{\wedge}$ and $\gamma_{\wedge}$ going to zero at large magnetization. For the wedge terms to be important, the plasma must be in a moderately magnetized regime ($\omega_e \tau_e \approx 1$).
	
	The error in the Epperlein \& Haines transport coefficients can alternatively be explained by their focus on the electric field. They did not consider the coefficients in terms of the induction equation, where the gradients of the coefficients are important. The fitting function in the Hall parameter that Epperlein \& Haines chose has a derivative that does not tend to zero as the magnetic field tends to zero, which is a physical requirement. The errors in the fits become more apparent when the coefficients are reformulated in a manner that more clearly demonstrates their physical effects in the induction equation.
	
	This paper demonstrates the importance of the new transport coefficients in laboratory plasmas, focusing on the thermally-driven cross-gradient-Nernst term. In section \ref{sec:underdense} the new coefficient is shown to reduce magnetic field twisting in under-dense systems relevant to MagLIF preheat \cite{gomez2020}. This setup was proposed in order to make the first measurements of the cross-gradient-Nernst transport \cite{walsh2020}. Section \ref{sec:premag} then shows how the cross-gradient-Nernst term can result in field twisting in pre-magnetized ICF capsules \cite{walsh2018a}; again, the new coefficients result in reduced twisting at low plasma magnetizations. However, the new coefficients do not universally reduce the impact of cross-gradient-Nernst transport. Section \ref{sec:selfgen} compares ICF hot-spot simulations without MHD, with MHD but no cross-gradient-Nernst, full extended-MHD with the Epperlein \& Haines coefficients and with the new coefficients. Cross-gradient-Nernst transport is shown to be important in hot-spot cooling, increasing the penetration depth of a cold spike. An example of how the new $\gamma_{\wedge}$ coefficient affects perturbation growth in direct-drive ablation fronts was already given in one of the original updated coefficients publications \cite{sadler2021}
	
	Simulations in this paper use the Gorgon extended-MHD code \cite{ciardi2007,chittenden2004,walsh2017}. The code can use either the old Epperlein \& Haines coefficients \cite{epperlein1986} or the updated coefficients (see the supplementary material of reference \cite{sadler2021}). The coefficients calculated by Sadler, Walsh \& Li are used in this paper, although the coefficients from Davies, Wen, Ji \& Held have been implemented and give no noticeable differences \cite{davies2021} (as expected from figure \ref{fig:gamma}). Epperlein \& Haines provided tabulated data for the coefficients at specific values of Z; these values are interpolated in the simulations. The new coefficients provide a polynomial fit to ionization, which is preferable for reducing computations and eliminating discontinuities when gradients are taken in Z.

	\section{Underdense Plasmas\label{sec:underdense}}
	
	This section investigates the impact of cross-gradient-Nernst on magnetized under-dense gases heated by a laser. This setup is relevant to MagLIF \cite{slutz2010} and mini-MagLIF preheat \cite{barnak2017}, but was also investigated as a means of measuring specific extended-MHD transport terms for the first time \cite{walsh2020}. Particularly relevant was the suggestion that quantification of a twisted magnetic field component  could be used to measure the cross-gradient-Nernst advection velocity \cite{walsh2020}. Simulations in that experiment design paper used the erroneous Epperlein \& Haines coefficients \cite{epperlein1986}, which are shown here to give excessive magnetic field twisting at low magnetization. 
	
	The configuration used as demonstration here is as follows. A low density ($5 \times 10^{19}$ atoms/cm$^3$) deuterium gas is irradiated by a $5J$ beam with a $0.5ns$ square temporal pulse. The beam has a Gaussian spatial profile with a standard deviation $\sigma=100\mu$m at best focus. The beam is tapered along its propagation direction such that at $r=\sigma$ the angle of the rays to the axis satisfies $\sin\theta = 1/12$. A 1T magnetic field is applied along the laser propagation axis. The beam tapering is purposefully used to induce magnetic field twisting by cross-gradient-Nernst transport \cite{walsh2020}. 
	
	2-D $\underline{r}$,$\underline{z}$ simulations are used, with the laser drive and applied magnetic field along $\underline{z}$. The laser is treated as individual rays that are traced through the plasma and deposit their energy by inverse Bremsstrahlung. A $460\mu$m axial extent is used, with radial and axial resolution of $1\mu$m.
	
	As the gas density is low, very little energy couples to the system. At $0.1$ns the total absorption along the $460\mu$m axial extent is 2.5\%, reducing to 1\% by the end of the laser pulse when the plasma temperature is higher ($>100$eV). As the Hall Parameter is low for the chosen setup ($\omega_e \tau_e < 0.2$ everywhere) thermal conduction effectively transports energy radially. Simultaneously, the Nernst term advects magnetic field away from the beam center, reducing the field strength to 0.3T by 0.5ns. The magnetic flux piles up at the thermal conduction front, peaking at 2T.
	
	The impact of cross-gradient-Nernst can be seen by decomposing the advection term in equation \ref{eq:mag_trans_new}:
	
	\begin{equation}
	\bigg[ \frac{\partial \underline{B}}{\partial t} \bigg] _{N \wedge} + (\underline{v}_{N\wedge} \cdot \nabla) \underline{B} =   -\underline{B}(\nabla \cdot \underline{v}_{N\wedge}) + (\underline{B} \cdot \nabla) \underline{v}_{N\wedge}  \label{eq:theory_velocity_B}
	\end{equation}
	
	As the temperature gradient is predominantly radial and the magnetic field is predominantly axial, the cross-gradient-Nernst velocity is purely in $\underline{\theta}$. Therefore, equation \ref{eq:theory_velocity_B} reduces to $\big[ \frac{\partial B_{\phi}}{\partial t} \big] _{N \wedge} = B_z \frac{\partial v_{N\wedge\theta}}{\partial z} $. This term represents magnetic field twisting. Following a single axial magnetic field line, if $\underline{v}_{N\wedge} = 0$ at the bottom and $\underline{v}_{N\wedge} \ne 0$ at the top, a $B_{\theta}$ field will be generated by twisting in between. By tapering the laser beam, this setup purposefully makes this happen; near the laser edge a magnetic field line will pass between an unheated and a heated regime, resulting in twisting of the magnetic field \cite{walsh2020}. This process can be viewed as a dynamo driven by electron heat-flow.
	
	Figure \ref{fig:underdense} compares the twisted magnetic field component at $t=1.0$ns using the old Epperlein \& Haines transport coefficients and the updated coefficients \cite{sadler2021,davies2021}. The Biermann Battery generation has been neglected in these simulations, which means that $B_{\theta}$ can only be generated by cross-gradient-Nernst twisting. As the experiment here is in the low magnetization regime, the difference in the $\gamma_{\wedge}$ coefficient is substantial; at the laser spot edge, where much of the twisting takes place, $\omega_e \tau_e \approx 0.05$, resulting in a factor of 4 reduction in the cross-gradient-Nernst coefficient. By 1.0ns the peak $B_{\theta} \approx$0.08T for the old coefficients, and $\approx$0.01T using the new coefficients. 
	
	 The new coefficients change the regime where significant $B_{\theta}$ is generated. Designs using the old coefficients aimed for the underdense plasma being as low magnetization as possible. Now, however, it is clear that the plasma should be balanced in the moderately magnetized regime, with the cross-gradient-Nernst velocity peaking at around $\omega_e \tau_e \approx 0.8$ for deuterium \cite{sadler2021}. This is still realizable, with the density, laser power and applied field all acting as free variables. An experiment with the plasma parameters defined here but scanning applied field strength is expected to observe the true $\gamma_{\wedge}$ dependence.

	However, with the effect of cross-gradient-Nernst reduced, the Biermann Battery contribution to $B_{\theta}$ becomes more important. At 0.1ns for the simulations shown here the $B_{\theta}$ field from Biermann Battery is of the same order as from cross-gradient-Nernst. Biermann requires a density gradient in the plasma, which only develops later in time, at time-scales on the order of the laser radius divided by the sound speed. Therefore, probing early in time before the plasma hydrodynamically expands is important to distinguish Biermann from cross-gradient-Nernst.
	
	\section{Pre-magnetized ICF Capsules\label{sec:premag}}
	
	\begin{figure}
		\centering
		\includegraphics[scale=0.6]{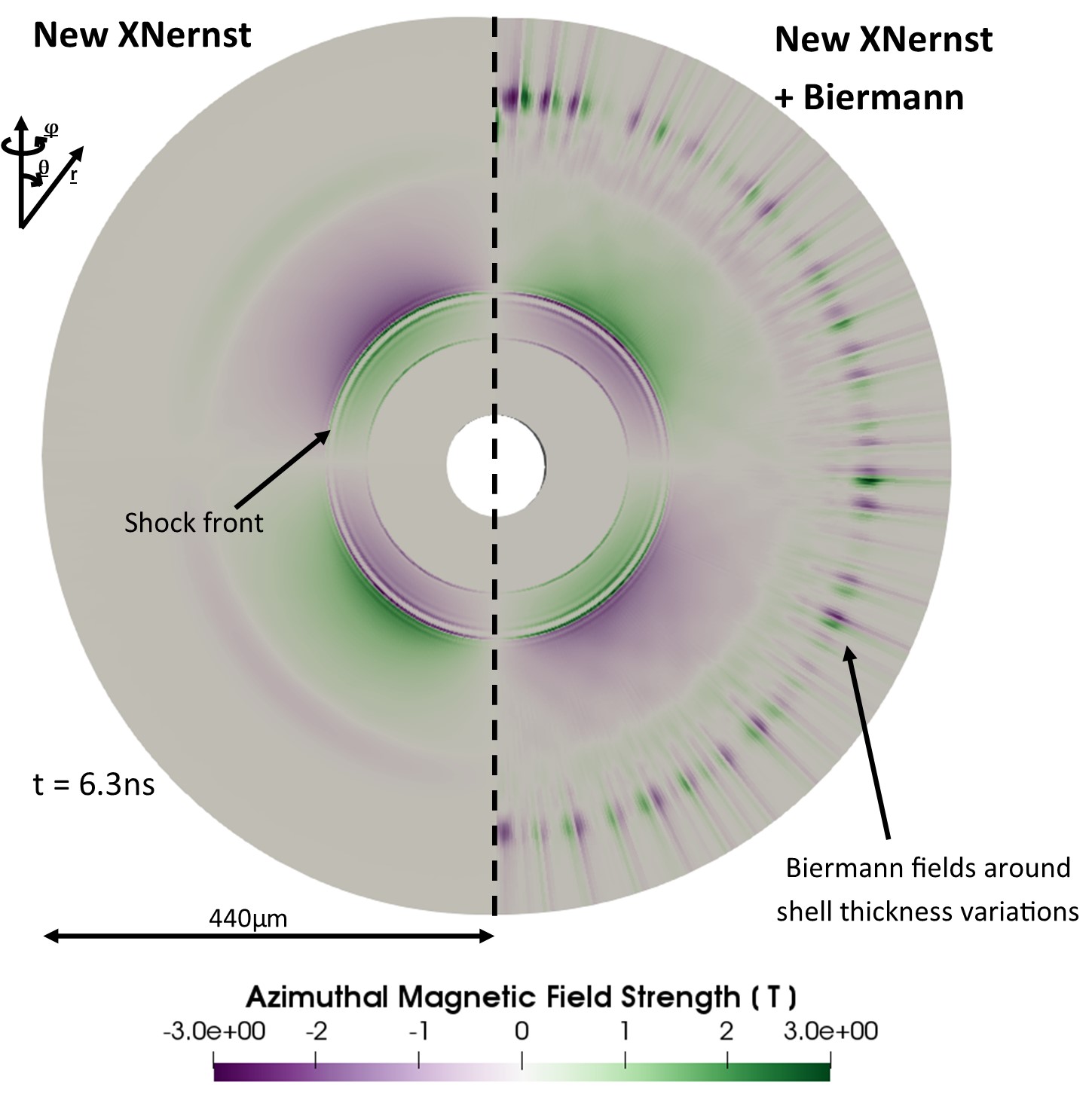}
		\caption{ \label{fig:Premag_XNernst_Biermann} $B_{\phi}$ at 6.3ns for a warm indirect-drive implosion with an applied axial field of 5T. Left is a simulation with cross-gradient-Nernst included (using the new coefficients), while right shows a case with both cross-gradient-Nernst and Biermann included. }
	\end{figure}

	Axial magnetic fields can be applied to ICF capsules to reduce thermal cooling of the fuel during stagnation \cite{chang2011,perkins2017,walsh2019}. Throughout the implosion the magnetic field compresses along with the plasma; this happens dominantly at the capsule waist, where the implosion velocity is perpendicular to the magnetic field lines. At the poles the magnetic field remains uncompressed, as the implosion velocity is along field lines. 
	
	It has been suggested that the cross-gradient-Nernst term should twist the applied magnetic field during the implosion \cite{davies2015}. Preliminary simulations showed significant twisting by bang-time, with the yield increased by as much as 10\% due to the extra path length required for heat to travel along field lines out of the hot-spot \cite{walsh2018a}. Again, these calculations used the old Epperlein \& Haines coefficients. Here, as with the twisting seen in the underdense configuration in section \ref{sec:underdense}, the twisting is found to be lower when the new transport coefficients are used. 
	
	The simulated configuration is a $D_2$ gas-filled HDC indirect-drive capsule that is used on the National Ignition Facility, although these results broadly apply to all magnetized spherical implosions. A 5T magnetic field is applied axially. $1\mu$m radial resolution is used throughout the implosion and 180 polar cells are used in the simulation range $\theta = 0,\pi$. Before the first shock converges on axis the simulations are transferred to a cylindrical mesh.
	
	Moderate HDC shell thickness variations are initialized \cite{casey2021}, such that the capsule is mildly perturbed by neutron bang-time. 1600 perturbations with random amplitude and mode are used, with each chosen in the range $\epsilon=0,0.5$nm and $k=0,180$ respectively.
	
	\begin{figure}
		\centering
		\includegraphics[scale=0.6]{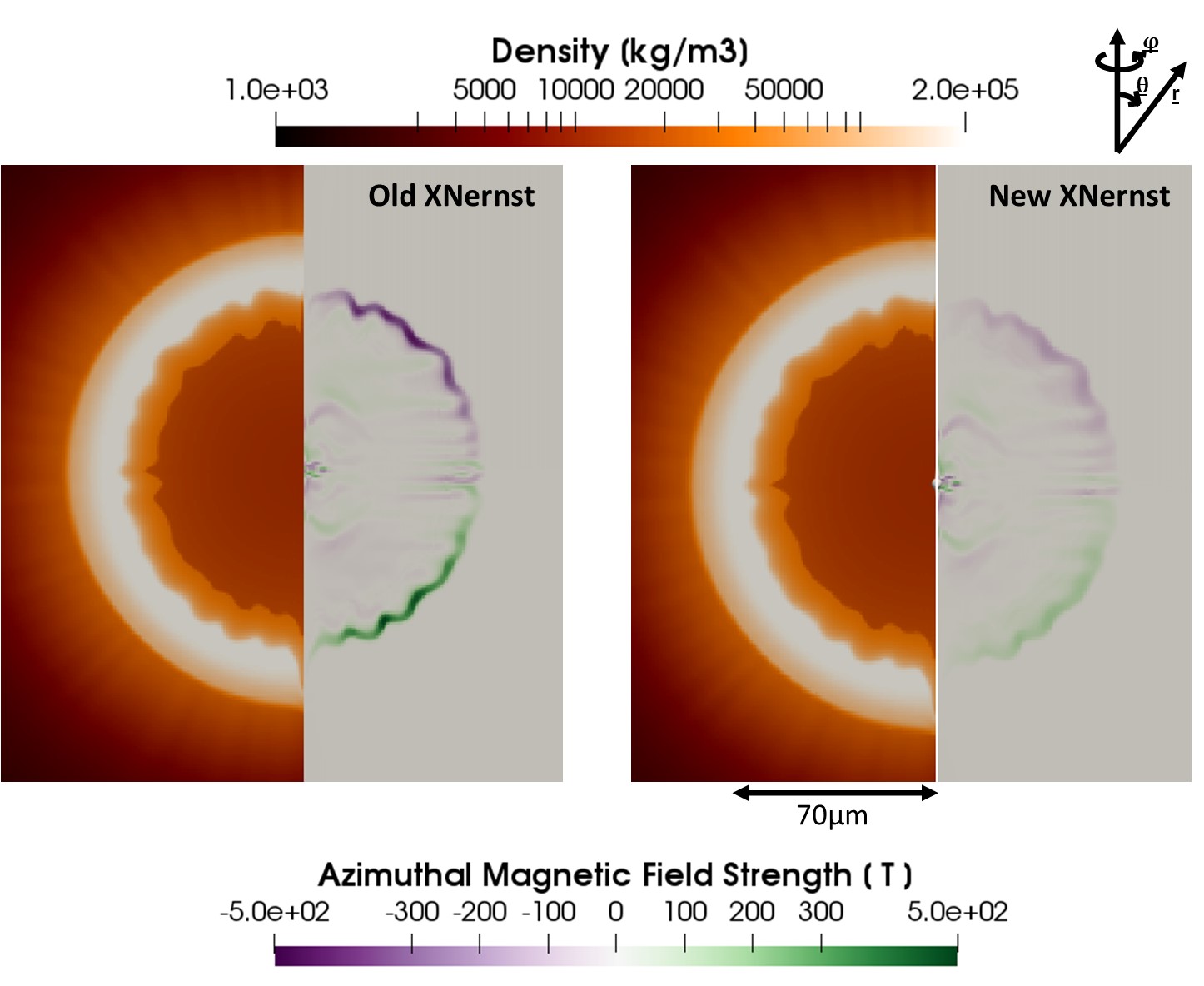}
		\caption{ \label{fig:Premag_XNernst} Density and $B_{\phi}$ at neutron bang-time ($t=7.3ns$) using the old Epperlein \& Haines coefficients (left) and the updated versions (right). }
	\end{figure}
	
	While for the underdense configuration a tapered beam was used to induce twisting, a spherical implosion naturally has twisting peaking at $\theta = \pi/4,3\pi/4$. This happens because the cross-gradient-Nernst velocity $- \underline{\hat{b}}\times \nabla T_e$ is maximum at the capsule waist and zero at the poles. Therefore, the axial magnetic field at the waist is displaced in $\phi$ but at the poles it is stationary; in between the waist and the poles the magnetic field lines are twisted out of the plane.
	
	Twisting of magnetic field lines introduces a closed field line component, $B_{\phi}$. This is of great interest to the magneto-inertial fusion community, as this component is effectively compressed during the implosion and results in additional thermal energy containment. Schemes have been considered to apply a purely $B_{\phi}$ component to a capsule \cite{hohenberger2011}, although none have been demonstrated for spherical implosions.
	
	The cross-gradient-Nernst velocity is significant both in the in-flight shock-compressed gas and in the stagnating hot-spot. For the configuration chosen here the in-flight phase is not changed significantly by the new coefficients, as the low density and high temperature gas is magnetized even with $B_0 = 5$T. For lower applied fields the coefficients may make a significant difference, but these regimes are not of interest for magneto-inertial fusion. 
	
	Figure \ref{fig:Premag_XNernst_Biermann} shows the $B_{\phi}$ field profile induced by the cross-gradient-Nernst effect at 6.3ns, which is before the first shock converges onto the axis. While $B_{\phi}$ only reaches 3T at this time, the field twisting is significant: $B_{\phi}/|\underline{B}|=1/3$ in locations near the shock-front. 
	
	Also shown in figure \ref{fig:Premag_XNernst_Biermann} is a simulation with the Biermann Battery self-generated magnetic fields included. This term requires an asymmetric implosion to be important, of which there are two sources in this implosion. First of all, the applied field results in anisotropy of the heat-flow; here that results in a hotter compressed gas at the waist compared with the poles. Also, the HDC thickness variations generate magnetic fields. The interest here is in the fuel $B_{\phi}$ component, which is dominated by cross-gradient-Nernst twisting for $B_0 = 5$T. Larger applied fields result in suppression of cross-gradient-Nernst and enhancement of the thermal conductivity anisotropy, enhancing the impact of Biermann. 
	
	Figure \ref{fig:Premag_XNernst} shows the density and out-of-plane magnetic field component at bang-time. Biermann Battery generation of magnetic field has been turned off in the code for these simulations, which means that all of $B_{\phi}$ is from cross-gradient-Nernst twisting. On the left is a case using the old cross-gradient-Nernst coefficients from Epperlein \& Haines, while on the right uses the updated coefficients. The new coefficients lower the peak $B_{\phi}$ from 500T to 250T. Nonetheless, the twisting is significant, with $B_{\phi}/|\underline{B}|$ up to 0.3 throughout the hot-spot.
	
	The twisting introduces several additional effects. Firstly, thermal conduction and Nernst are moderately reduced, as these components become partially out of the simulation plane; Cross-gradient-Nernst and Righi-Leduc then have components within the simulation plane. If the magnetic tension becomes significant then the twisted field would also be expected to induce plasma motion in $\phi$; however, as cross-gradient-Nernst is suppressed for the large applied fields required for tension to be important \cite{perkins2017,walsh2019}, this may not be a realizable regime. For $B_0 = 5$T the plasma $\beta$ is too large for significant motion.
	
	
	While the simulations here focused on indirect-drive implosions, cross-gradient-Nernst also twists magnetic fields in pre-magnetized direct-drive ablation fronts \cite{walsh2020a}, where the temperature gradients drive extreme heat-flows. However, the $B_{\phi}$ component is effectively advected by the plasma and Nernst velocities, lowering its impact.
	
	\section{Self-Generated Fields in ICF Hot-spots\label{sec:selfgen}}
	
	\begin{figure}
		\centering
		\includegraphics[scale=0.5]{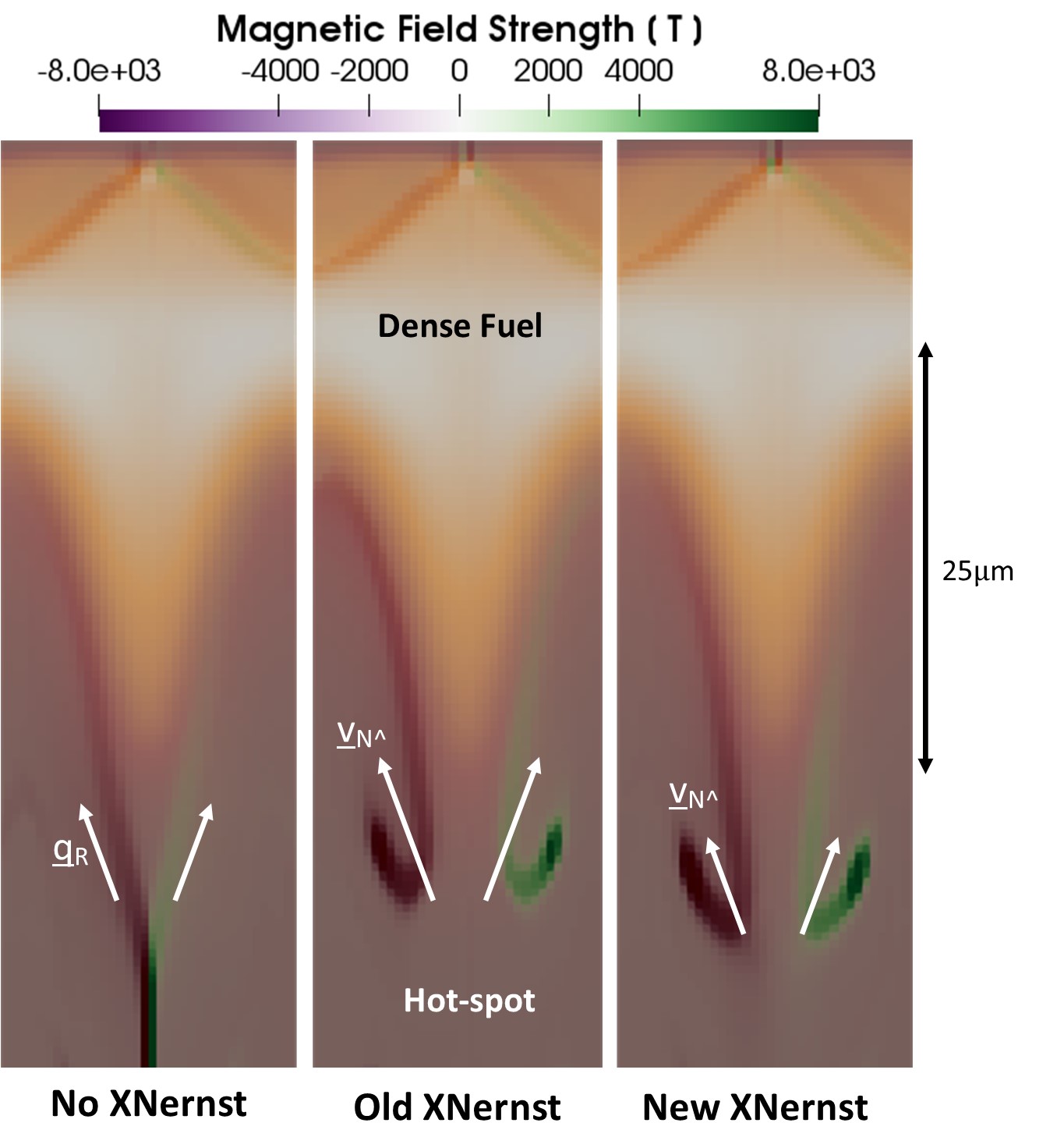}
		\caption{ \label{fig:ss_B} Self-generated magnetic fields at bang-time around a cold spike pushing into a hot-spot. 3 cases are shown. Left: a simulation without cross-gradient-Nernst advection included. Middle: a simulation using the Epperlein \& Haines transport coefficients \cite{epperlein1986}. Right: a simulation using the updated transport coefficients \cite{sadler2021,davies2021}.}
	\end{figure}
	
	This section looks into the impact of cross-gradient-Nernst on self-generated magnetic field profiles in regular ICF hot-spots, where magnetic fields have not been externally applied. 
	
	Magnetic fields are generated by the Biermann Battery mechanism around perturbations, with estimated field strengths up to 10kT \cite{walsh2017}. These fields are predominantly generated in the stagnation phase, when the temperature and density gradients are largest \cite{walsh2021}. Recent advances have been made in the theoretical understanding of magnetic flux generation in these systems, with more flux being generated around high mode and large amplitude perturbations \cite{walsh2021}.
	
	While it was expected that self-generated magnetic fields would reduce heat loss from ICF hot-spots,  research found that magnetization of the electron population introduced Righi-Leduc heat-flow, which enhanced cooling \cite{walsh2017}. As cross-gradient-Nernst is the magnetic transport analogue of Righi-Leduc heat-flow, it is important in these systems. 
	
	The simulations here use the indirect-drive HDC design N170601 is used, although the physics is also applicable to direct-drive hot-spots. An isolated 200nm HDC shell thickness variation is imposed at the capsule pole, causing a cold spike to push into the hot-spot. Shell thickness variations have been found to be a significant degradation mechanism for HDC implosions \cite{casey2021}; the variation applied here is not based on any target fabrication specifications, and is instead used as a demonstration of magnetized heat-flow in ICF implosions. 
	
	The simulations here are 2-D and are in spherical geometry until the first shock converges onto the axis ($t\approx 7.6$ns). The resolution up to this time is $1\mu$m radially, with 180 cells in the polar direction from $\theta = 0,\pi$. The simulations are then remapped into cylindrical geometry for the stagnation phase, when the resolution is $\frac{1}{2}\mu$m.
	
	Figure \ref{fig:ss_B} shows the magnetic field distribution at neutron bang-time ($t= 8.5$ns) for simulations with different cross-gradient-Nernst physics included. The magnetic field profiles are in green (into the page) and purple (out of the page), plotted over the density so that the proximity to the spike can be seen. The dense fuel is at the top of the figure and the hot-spot at the bottom.
	
	\begin{figure}
		\centering
		\includegraphics[width=0.5\textwidth]{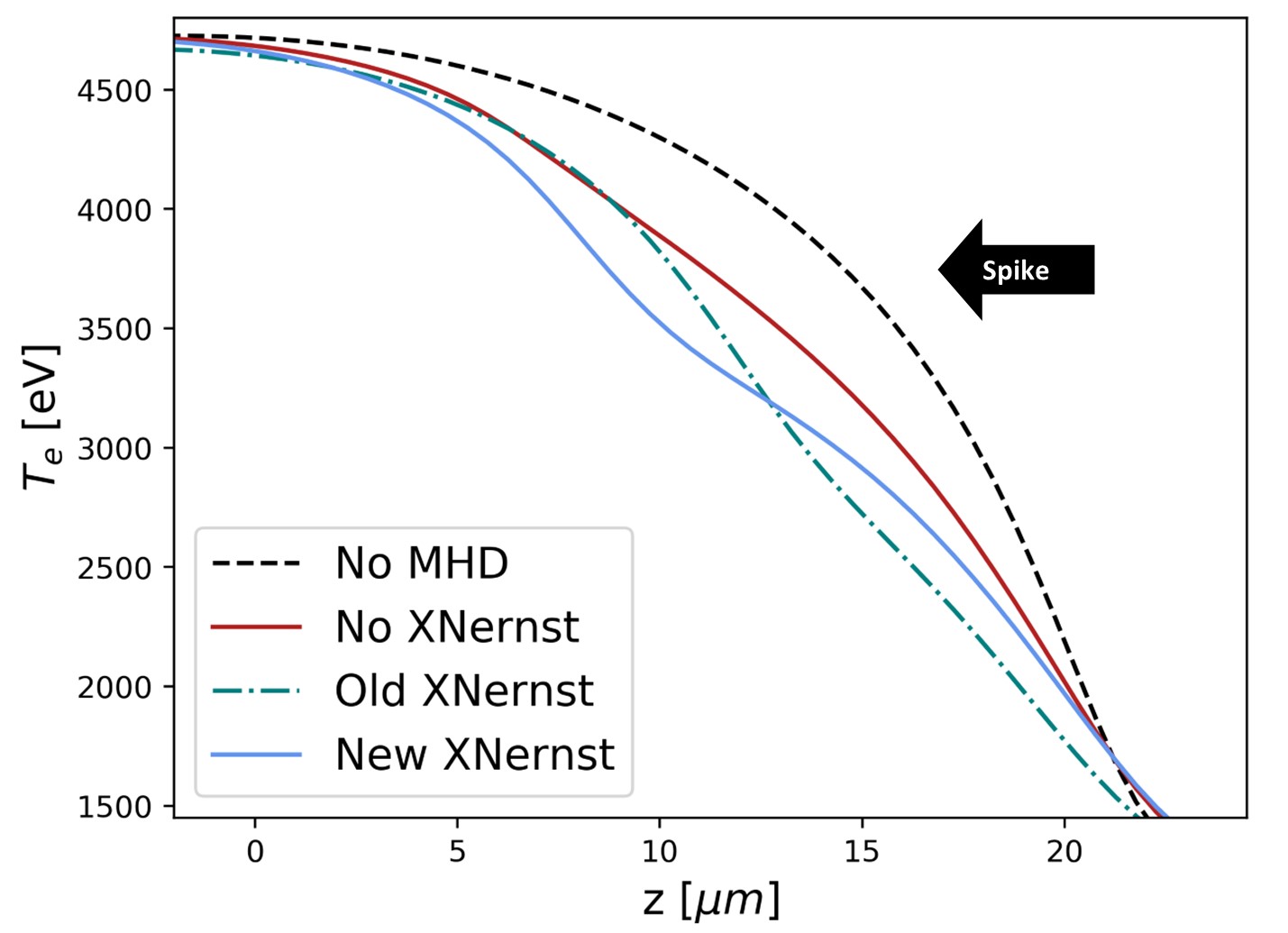}
		\caption{ \label{fig:ss_Te} Electron temperature along the axis of a hot-spot spike at bang-time. Temperature profiles are shown for cases: without any self-generated fields included; with extended-MHD but no cross-gradient-Nernst advection; using the Epperlein \& Haines transport coefficients \cite{epperlein1986}; using the updated coefficients \cite{sadler2021,davies2021}.}
	\end{figure}
	
	First in figure \ref{fig:ss_B} is a case with no cross-gradient-Nernst advection. The Righi-Leduc heat-flow direction is shown with white arrows. Righi-Leduc acts to cool the spike tip, which results in regular Nernst compressing the field onto the simulation axis. This process causes numerical issues, advecting all of the magnetic field into a single line of cells. Once the cross-gradient-Nernst is included, the magnetic field moves in the same direction as Righi-Leduc, preventing the field profile from compressing onto the axis. It can be seen in figure \ref{fig:ss_B} that the new cross-gradient-Nernst coefficient does not advect as rapidly as the Epperlein \& Haines version. 
	
	It is posited that the updated coefficients will always be important in systems with self-generated magnetic fields and thermal conduction; there is always a null-point of magnetic field, resulting in no electron magnetization. Even if the peak magnetization is large enough to be in a regime where the Epperlein \& Haines coefficients are valid, the magnetization will decrease to zero in the surrounding regions, passing through the regime where the old coefficients are invalid.

	Figure \ref{fig:ss_Te} shows the impact of the imposed spike on the hot-spot temperature for simulations with various MHD packages included. The electron temperature is plotted along the spike axis. Without any MHD included the spike does not propagate as far; this suggests that current design calculations of ICF implosions (both directly and indirectly driven) are underestimating the impact of perturbations. 
	
	Simulations with the new coefficients included result in the greatest discrepancy with the estimates that do not include Biermann generation; at the $T_e = 3.5$keV contour the spike has penetrated $7 \mu$m further due to the magnetization of heat-flow. The simulations without cross-gradient-Nernst are less affected by the magnetic fields, as the fields are transported by the standard Nernst term into a one-cell width (see figure \ref{fig:ss_B}). 
	
	The new coefficients give a greater impact of MHD on spike penetration compared with the old coefficients. This can be understood from the fact that the new coefficients result in the magnetic field moving with the Righi-Leduc heat-flow. In contrast, the old coefficients allowed the magnetic flux to move ahead of the Righi-Leduc heat-flow into regions where it would have a lower impact on the plasma magnetization. This can be seen in figure \ref{fig:coeff}, which shows that the ratio of Righi-Leduc coefficient to cross-gradient-Nernst coefficient decreasing to zero for low magnetization.

	Closer analysis of the impact of MHD on capsule perturbation growth for a variety of mode numbers and amplitudes will be the subject of a future publication.

	\section{Conclusions}
	
	In summary, updated magnetic transport coefficients that accurately replicate kinetic simulations at low electron magnetizations \cite{sadler2021,davies2021} have been shown to be important across a range of laboratory plasma conditions. These coefficients are the new standard for implementation into extended-MHD codes. While the two references give different fits for the coefficients, they are found to be practically equivalent.
	
	With an external magnetic field applied, the cross-gradient-Nernst term tends to twist the magnetic field \cite{davies2017,walsh2018a}; using the new coefficients reduces twisting in the low magnetization ($\omega_e \tau_e < 1$) regime. This result impacts attempts to measure cross-gradient-Nernst for the first time \cite{walsh2020} as well as the design of pre-magnetized capsule implosions \cite{chang2011,perkins2017,walsh2018a,walsh2019}. Twisting is still possible in the moderate magnetization regime ($\omega_e \tau_e \approx 1$).
	
	For systems with significant self-generated magnetic fields \cite{farmer2017,walsh2017,sadler2020a} the new coefficients result in the magnetic fields moving with the Righi-Leduc heat-flow. This is found to enhance the impact of Righi-Leduc, as shown in direct-drive ablation fronts (where Righi-Leduc reduces perturbation growth) \cite{sadler2021} and in ICF hot-spots (where Righi-Leduc enhances perturbation growth).
	
	In addition to being more physically accurate, the new coefficients have been found to increase numerical stability, as they introduce fewer discontinuities into the simulations.
	
	\section*{Acknowledgements}
	This work was performed under the auspices of the U.S. Department of Energy by Lawrence Livermore National Laboratory under Contract DE-AC52-07NA27344 and by the LLNL-LDRD program under Project Number 20-SI-002. The simulation results were obtained using the Imperial College High Performance Computer Cx1..
	
	This document was prepared as an account of work sponsored by an agency of the United States government. Neither the United States government nor Lawrence Livermore National Security, LLC, nor any of their employees makes any warranty, expressed or implied, or assumes any legal liability or responsibility for the accuracy, completeness, or usefulness of any information, apparatus, product, or process disclosed, or represents that its use would not infringe privately owned rights. Reference herein to any specific commercial product, process, or service by trade name, trademark, manufacturer, or otherwise does not necessarily constitute or imply its endorsement, recommendation, or favoring by the United States government or Lawrence Livermore National Security, LLC. The views and opinions of authors expressed herein do not necessarily state or reflect those of the United States government or Lawrence Livermore National Security, LLC, and shall not be used for advertising or product endorsement purposes.

	Research presented in this article was also supported by the Laboratory Directed Research and Development program of Los Alamos National Laboratory, under the Center for Nonlinear Studies project number 20190496CR. This research was supported by the Los Alamos National Laboratory (LANL) through its Center for Space and Earth Science (CSES). CSES is funded by LANL’s Laboratory Directed Research and Development (LDRD) program under project number 20180475DR.
	
	Finally, the information, data, or work presented herein was funded in part by the Advanced Research Projects Agency-Energy (ARPAE), U.S. Department of Energy, under Award No. DE-AR0001272, by the Department of Energy Office of Science, under Award No.
	DE-FG02-04ER54746, by the Department of Energy National	Nuclear Security Administration under Award No. DENA0003856, the University of Rochester, and the New York State	Energy Research and Development Authority.
		
	\section*{References}
	
	\bibliographystyle{plainnat}
	\ifdefined\DeclarePrefChars\DeclarePrefChars{'’-}\else\fi

\end{document}